\documentclass[aps, prd, twocolumn, superscriptaddress, longbibliography]{revtex4-2}
\usepackage{amsmath}
\usepackage{amssymb} 
\usepackage{graphicx}
\usepackage{enumerate}
\usepackage{color}
\usepackage{tensor}
\usepackage{mathrsfs}
\usepackage{hyperref}
\usepackage[normalem]{ulem}
\usepackage[dvipsnames]{xcolor}
\usepackage{accents}
\usepackage{scalerel}
\usepackage{bm}
\usepackage{anysize}
\usepackage{siunitx}

\usepackage{multirow}

\makeatletter
\newcommand*{\leqdef}{\mathrel{\rlap{%
			\raisebox{0.25ex}{$\m@th\cdot$}}%
		\raisebox{-0.25ex}{$\m@th\cdot$}}%
	=}
\newcommand*{\reqdef}{=\mathrel{\rlap{%
			\raisebox{0.25ex}{$\m@th\cdot$}}%
		\raisebox{-0.25ex}{$\m@th\cdot$}}
}
\makeatother

\begin{document}
\title{Gravitational wave propagation in $f(R)$ models: New parametrizations and observational constraints}
\author{Isabela S. Matos}
\email{isa@if.ufrj.br}

\author{Maur\'\i cio O. Calv\~ao}
\email{orca@if.ufrj.br}

\author{Ioav Waga}
\email{ioav@if.ufrj.br}

\affiliation{Universidade Federal do Rio de Janeiro,
	Instituto de F\'\i sica, \\
	CEP 21941-972 Rio de Janeiro, RJ, Brazil}

\begin{abstract}
	Modified gravity (MG) theories predict, in general, that the ratio of gravitational wave (GW) to electromagnetic (EM) luminosity distances, $\Xi$, differs from its general relativity (GR) value of unity at cosmological scales, thus providing another perturbative probe to MG. In this paper, we introduce new phenomenological parametrizations for both the Friedmann-Lemaître-Robertson-Walker (FLRW) background evolution of $f(R)$ models, via the dark energy equation of state parameter, $w_{\text{DE}}$, and for $\Xi$ in this class of theories. We simulate a mock dataset for the Einstein Telescope (ET) of 1000 GW signals from binary neutron star (BNS) mergers and redshift information from their EM counterpart, exploring the consequent constraints on the relevant gravitational, cosmological and phenomenological parameters. As a model of particular interest, we take $\gamma$-gravity theory and investigate whether it could be distinguished from GR's $\Lambda$CDM model. We then combine our results with actual data from type Ia supernovae (SNIa) and combined baryon
	acoustic oscillations (BAO) and cosmic microwave background (CMB) observations. Additionally, we also investigate the potential bounds to $f_{R0}$ for any viable $f(R)$ whose background evolution is indistinguishable from the standard model of cosmology above a certain redshift, showing that, for a $\Lambda$CDM fiducial model, ET data would provide $|f_{R0}|<10^{-2}$ at a 95\% level. We conclude altogether that probing the redshift evolution of the GW luminosity distance from detections of the ET in its first running decade will not substantially help constraining $f(R)$ theories of gravity.
\end{abstract}

\maketitle
\section{Introduction}
\label{sec:intro}

The ground-breaking first direct detection of gravitational waves (GWs) \cite{Abbott2016a} ushered in a much anticipated era not only for astronomy and cosmology but also for general physics. Their generation and propagation probe the strong field regime and cosmological scales, making it possible, \textit{inter alia}, to test general relativity (GR) against modified gravity (MG) theories. The plethora of MG candidates is astonishing, and the exploration of their theoretical and observational viability, in general and particularly for cosmology, is of paramount importance \cite{Clifton2012,Ferreira2019,Heisenberg2019,Ishak2019,Amendola2020}. In this paper, we will be concerned with cosmological constraints on the so-called $f(R)$ MG \cite{DeFelice2010,Soutiriou2010}, characterized by the following action:
\begin{align}
	S = \int [R + f(R)]\sqrt{- g}\,d^4x + S_m\,.
\end{align}
As is well known, these theories have been explored more recently since they can modify the cosmic evolution at late times and explain cosmic acceleration. Here, particular attention will be given to $\gamma$ gravity \cite{ODwyer2013}.

More broadly, in a Friedmann-Lemaître-Robertson-Walker (FLRW) background metric, for a variety of MG theories including GR, there are 2 independent degrees of freedom for a tensor perturbation in the TT gauge \cite{Hwang2001, Nishizawa2018}. Writing the perturbed field equations, one may derive several modifications to the propagation of these GW modes when compared to GR's prediction, as discussed in \cite{Saltas2014, Hwang1991, Tsujikawa2014, Belgacem2018a}. Among them, there might be a GW speed of propagation $c_T$, not necessarily equal to that of light, which was, however, recently constrained to obey $|c_T - 1| < 10^{-15}$ at low redshifts \cite{Abbott2017c}. We will, therefore, throughout the paper, set $c_T = c = 1$. Without the addition of any anisotropic stress and in the context of $f(R)$ theories, those effects, as will be seen, besides the change in the background scale factor as
compared, for instance, to $\Lambda$CDM, reduce to a modification in the friction term of the equation describing GW propagation.

Furthermore, as discussed in the pioneering work \cite{Schutz1986}, the detection of GWs emitted by binary sources provide the inference of a (properly defined) GW luminosity distance $\mathcal{D}^{\,gw}_L$, ascribed to the source at emission through its measured signal. It differs from the usual electromagnetic (EM) luminosity distance $\mathcal{D}^{\,em}_L$ in MG due to the friction term in our context. However, even within the scope of GR, when $\mathcal{D}^{\,gw}_L = \mathcal{D}^{\,em}_L$,  it is possible to investigate additional constraints, for instance, to the dark energy equation of state \cite{Cai2017, Zhang2019} or the total neutrino mass \cite{Wang2018} via GW detections, provided the redshifts of the sources are also known. A catalog of GWs from cosmologically distant binary sources with EM counterpart, e.g., $\gamma$ ray bursts, is, therefore, a probe for deviations from $\Lambda$CDM. 

In this work, we will study the potentiality of the Einstein Telescope (ET) project to provide constraints to $f(R)$-like theories of gravity and the cosmological parameters via several observations of GWs from binary neutron star (BNS) mergers plus detections of the corresponding short $\gamma$-ray bursts \cite{DAgostino2019, Belgacem2018a}. ET is a proposed underground cryogenic third generation GW observatory, which is expected to see, in a possibly optimistic estimate, about 1000 of such multimessenger events, up to a redshift $z = 2$, within a decade of running \cite{Maggiore2020, Sathyaprakash2010}. Signals from black hole-neutron star binaries, which are expected to be much more scarce \cite{ArcaSedda2020}, were not included in this work.

The structure of the paper follows. In Sec. \ref{sec:GW_in_MG}, we briefly introduce how GW propagation in spatially flat FLRW spacetime is modified in $f(R)$-like theories, presenting the expression of the GW luminosity distance for this family of models and discussing its asymptotic regime. In Sec. \ref{sec:bayes_inference}, we write the distributions that will be used in our simulations, presenting the waveform and the errors considered for the distance, and then derive the posterior probability. In Sec. \ref{sec:gamma_gravity}, we discuss the main features of the $\gamma$-gravity $f(R)$ theory, and in Sec. \ref{sec:parametrizations}, we propose new independent parametrizations for the dark energy equation of state parameter and for the ratio of luminosity distances, exemplifying how well they recast the corresponding behaviors in Hu-Sawicki and $\gamma$-gravity theories. Finally, in Sec. \ref{sec:results}, we present the results of our simulations of GW detections by ET for three families of models: $\Lambda$CDM, $f(R)$ in the asymptotic regime, and the phenomenological set of models described by the parametrizations. We also combine our constraints with the ones from type Ia supernovae (SNIa) and baryon acoustic oscillations/cosmic microwave background (BAO/CMB). In Sec. \ref{sec:discussion}, we discuss our results and future perspectives.

\section{Gravitational wave propagation in alternative theories of gravity} \label{sec:GW_in_MG}

\noindent 

We consider here two ingredients introduced by MG theories in the propagation of GWs through cosmological distances: a modified friction term, quantified by a function called $\delta$ that vanishes in GR, and the modification in the background evolution. In this context, and with vanishing three-curvature, the Fourier transforms of the two independent metric perturbation components $h_P\; (P = +, \times)$ evolve according to \cite{Hwang2001}
\begin{align}
h_P''(\eta, \bm{k}) + 2\mathcal{H}(\eta)[1 - \delta(\eta)]h_P'(\eta, \bm{k}) & \nonumber \\ + \, k^2 h_P(\eta, \bm{k}) & = 0\,,	\label{delta_propagation}
\end{align}
where $\eta$ is the conformal time, $' \leqdef d/d\eta$, and $\mathcal{H} \leqdef a'/a$ is the conformal Hubble parameter.
By implicitly defining an effective scale factor $\tilde{a}$,
\begin{equation}
\frac{\tilde{a}'}{\tilde{a}} \leqdef (1 - \delta)\mathcal{H}\,,	\label{eff_scale_factor}
\end{equation}
one can show \cite{Belgacem2018a} that the solution to Eq.~(\ref{delta_propagation}) at scales such that $k^2 \gg \tilde{a}''/\tilde{a}$ has a time dependence of
\begin{equation}
h_P(\eta) \sim \frac{\sin(k\eta + \varphi_P)}{\tilde{a}(\eta)}\,. \label{evolution_h}
\end{equation}
Solving Eq.~(\ref{eff_scale_factor}) for $\tilde{a}$ and choosing it to coincide with $a$ today ($z = 0$), we obtain
\begin{equation}
\frac{\tilde{a}}{a} = \exp\bigg[\int_0^z \frac{\delta(\bar{z})}{1 + \bar{z}}d\bar{z}\bigg]\,.
\end{equation}

It can be shown \cite{Maggiore2007} that at observation, when the source is a coalescent binary system and the wave propagates in a GR FLRW background, $h_P$ can be expressed as
\begin{align}
h^{\text{GR}}_P(\eta_{\textrm o}) = \frac{\mathcal{A}_P}{\mathcal{D}^{\,gw}_L(z_{\textrm e})}\,, \label{GR_wave}
\end{align}
where the subindices ‘o’ and ‘e’ stand for evaluation at observation and emission, respectively, $\mathcal{A}_P$ is a source-dependent amplitude, and $\mathcal{D}^{\,gw}_L$ is the usual GR EM luminosity distance $\mathcal{D}^{\,em}_L$. As long as the MG theory and GR approximately coincide near the source where the curvature is high, this GR result could be analogously obtained in alternative theories (with the corresponding scale factor) if we ignore the modification in the propagation due to $\delta$ so that the emitted waves are equal functions of the several parameters that characterize the source, at least up to some order of post-Newtonian (PN) expansion. Since we have to account for the modified propagation in Eq.~(\ref{evolution_h}), the additional factor $\tilde{a}/a$ appears, and thus,
\begin{align}
h_P(\eta_{\textrm o}) = \frac{\mathcal{A}_P}{\mathcal{D}^{\,em}_L(z_{\textrm e})} \exp\bigg[\int_0^{z_{\textrm e}} \frac{\delta(\bar{z})}{1 + \bar{z}}d\bar{z}\bigg]\,.
\end{align}
In order to mimic the form of Eq.~(\ref{GR_wave}) in MG, one defines the GW luminosity distance $\mathcal{D}^{\,gw}_L$ to be
\begin{align}
\mathcal{D}^{\,gw}_L(z_{\textrm e}) \leqdef \mathcal{D}^{\,em}_L(z_{\textrm e})\exp\bigg[-\int_0^{z_{\textrm e}} \frac{\delta(\bar{z})}{1 + \bar{z}}d\bar{z}\bigg]\,. \label{ratio_dist}
\end{align}

In the context of $f(R)$ theories, it is possible to show \cite{Hwang2001} that GWs do evolve in a FLRW background according to Eq.~(\ref{delta_propagation}), with 
\begin{align}
\delta = - \frac{f'_{R}}{2\mathcal{H}(1 + f_R)}\,,
\end{align}
where $f_R \leqdef df/dR$, and all the factors are evaluated at the background spacetime. Thus, it is straightforward to compute the ratio of luminosity distances in Eq.~(\ref{ratio_dist}),
\begin{align}
\frac{\mathcal{D}^{\,gw}_L(z)}{\mathcal{D}^{\,em}_L(z)} = \sqrt{\frac{1 + f_{R0}}{1 + f_R(z)}}\,,\label{ratio_dist_f(R)}
\end{align}
where the subindex 0 stands for evaluation ``today`''.

It is well known that $f(R)$ viable models can not currently be distinguished from GR's $\Lambda$CDM at the background level \cite{dosSantos2016}. Further, in these models, both $f_R(z) \sim 0$ and $H \sim H_{\Lambda\text{CDM}}$ above a certain redshift. On the other hand, constraints from large scale structure imply $f_{R0}$ to be small (see works \cite{Jain2013, Hu2007}, which reach the stringent bound $f_{R0} < 10^{-6}$). Taking these considerations as reasonable assumptions to be imposed for viable $f(R)$ theories, one may approximate Eq.~(\ref{ratio_dist_f(R)}) to
\begin{align}
\frac{\mathcal{D}^{\,gw}_L(z)}{\mathcal{D}^{\,em}_L(z)} = 1 + \frac{f_{R0}}{2}\,, \label{asymptotic_lum_dist}
\end{align}
whenever the binary sources are located at sufficiently high redshifts for the particular model of interest.

In light of the usual difficulty in computing the background evolution in $f(R)$ MG, the above approximation introduces a way in which the detection of GWs from distant sources may independently constrain $f_{R0}$ for a large family of viable models at once. However, we note that, even at the asymptotic regime, we expect that the luminosity distances will not differ dramatically. This approximate scheme will be considered only in Sec. \ref{subsec:fR0}, while a more complete treatment will be held with the help of parametrizations in the main simulations, as discussed later on.

\section{Bayesian inference with standard sirens} \label{sec:bayes_inference}

\subsection{Waveform and errors}

The waveform at detection, emitted at redshift $z$ by a binary neutron star (BNS) merger, deduced up to the third PN correction \cite{Blanchet2014} is given by:
\begin{align}
& \tilde{h}(f, z, D^{\,gw}_{L}, \bm{s}) = \mathcal{A} \mathcal{Q} \bigg(\frac{c}{f^{7}}\bigg)^{1/6}e^{i\Phi(f)}\,, \\
& \mathcal{A} \leqdef \sqrt{\frac{5}{96}}\frac{(GM_{c}/c^2)^{5/6}}{\pi^{2/3}D^{\,gw}_{L}}\sum^{6}_{j = 0}A_j(\eta) \bigg(\frac{\pi GMf}{c^3}\bigg)^{\frac{j}{3}}\,, \\
& \mathcal{Q} \leqdef \sqrt{F_+^2 (1 + \cos^2\iota)^2 + 4F_{\times}^2\cos^2\iota}\,,
\end{align}
where
\begin{align}
F_+ \leqdef  \frac{\sqrt{3}}{2}\bigg[\frac{1}{2}(1 + \cos^2\theta) \cos(2\phi)\cos(2\psi) & \nonumber \\ - \cos\theta \sin(2\phi)\sin(2\psi)\bigg]\,, & \nonumber \\
F_{\times} \leqdef  \frac{\sqrt{3}}{2}\bigg[\frac{1}{2}(1 + \cos^2\theta) \cos(2\phi) \sin(2\psi) & \nonumber \\ + \cos\theta \sin(2\phi)\cos(2\psi)\bigg]\,, &
\end{align}
are the antenna pattern functions \cite{Zhao2011} when the angle between arms equals $\pi/3$. Descriptions of the parameters of the source $\bm{s} \leqdef (m_1, m_2, \iota, \theta, \phi, \psi)$ appearing in the above expressions are presented in Table \ref{tab:variables}. The coefficients $A_j$ of the PN expansion can be found in \cite{DAgostino2019}, where we make the spins of the binary components vanish, for simplicity. The function $\Phi$ is the phase of the wave and depends on some of the mentioned and also other parameters. They were omitted since the phase is unimportant to our further calculations.

\begin{table*}[t] 
	\centering
	\caption{Variables}
	\label{tab:variables}
	\setlength{\extrarowheight}{5pt}
	
	\begin{tabular}{|c|c|c|}
		\hline
		Symbol & Description & In simulations \\
		\hline                  $\bm{\Theta}$ & Cosmological and gravitational parameters & $(\Omega_{m0}, H_0, A, z_t, z_f, \Xi_0, \nu)$ or only $f_{R0}$\\
		$\bm{s}$ & Parameters of the source & $(m_1, m_2, \iota, \theta, \phi, \psi)$ \\     
		$z$ & Redshift of the source at emission & Sampled from distribution (\ref{z_pdf}) up to $z = 2$\\
		$D^{\,gw}_{L}$ & GW luminosity distance of the source & Sampled from distribution (\ref{dl_prob})\\
		\multirow{2}{*}{$\mathcal{D}^{\,gw}_L(z, \bm{\Theta})$} & Theoretical prediction for the GW & Phenomenologically parametrized, Eq. (\ref{dl_parametrized}) \\ & luminosity distance &  [in Sec. (\ref{subsec:fR0}), Eq. (\ref{asymptotic_lum_dist})]\\
		$m_1, m_2$ & Masses of the binary components & Uniformly sampled from [1, 2] $M_{\odot}$ \\
		$\eta$ & Symmetric mass ratio $= m_1 m_2/M^2$ & - \\
		$M_c$ & Redshifted chirp mass $(1 + z)\eta^{\frac{3}{2}}$ & - \\
		$\iota$ & Angle of orbital inclination & Uniformly sampled from $[\ang{0}, \,\ang{20}]$ \cite{Zhao2011}\\
		$\theta, \phi$ & Direction of the line of sight & Uniform sampling in the sky\\
		$\psi$ & GW polarization angle  & Uniformly sampled from [0, 2$\pi$]\\
		\hline
	\end{tabular}
\end{table*}

In the Fisher matrix approximation for the parameter estimation from data of GW interferometers \cite{Maggiore2007}, if the luminosity distance can be considered independent of the remaining parameters, the signal to noise ratio given by
\begin{align}
\left[\text{SNR}(z, \bm{s})\right]^2 & \leqdef 4 \int_{f_{\text{low}}}^{f_{\text{up}}} \frac{\left|\tilde{h}(f, z, \mathcal{D}^{\,gw}_{L}(z), \bm{s})\right|^2}{S_n(f)}df\,,
\end{align}
is the relevant quantity to compute the errors in the inferred $D^{\,gw}_L$ (for a comprehensive discussion, see \cite{Cutler1994}). Here, $S_n$ is the power spectral density (PSD) of the detector's noise which, for ET, is expected to be typically the expression found in \cite{Zhao2011}, used hereafter, and $(f_{\text{low}}, f_{\text{up}}) = (1, 10^4)$ are the extremities of the interval, out of which, the noise is effectively infinite. Since there is a well-known degeneracy between the distance and the orbital angle of the source, we add a factor of 2 which roughly takes this into account \cite{DAgostino2019}, ending up with an instrumental error equaling
\begin{align}
\sigma_{\text{ins}}(z, \bm{s}) & = \frac{2 \mathcal{D}^{\,gw}_L(z)}{\text{SNR}(z, \bm{s})}\,. \label{sigma_ins}
\end{align}
Another important source of error is weak lensing, which, as in \cite{Sathyaprakash2010}, we assume  to be well modeled by
\begin{align}
\sigma_{\text{lens}}(z) = 0.05 z \mathcal{D}^{\,gw}_L(z) \label{sigma_lens}
\end{align}
and quadratically sum Eqs.~(\ref{sigma_ins}) and (\ref{sigma_lens}) to give the total error in each of the $N_\text{obs}$ estimated GW luminosity distances,
\begin{align} \label{sigma}
\sigma_i(z_i, \bm{s}_i) \leqdef \sqrt{\sigma_{\text{ins}}(z_i, \bm{s}_i)^2 + \sigma_{\text{lens}}(z_i)^2}\,,
\end{align}
for $ i = 1, ..., N_{\text{obs}}$. Here, we omit any dependence of the functions, e.g., the luminosity distance, with the gravitational or cosmological parameters $\bm{\Theta}$ since the errors will always be evaluated at the fiducial model employed in the simulations.

Other sources of uncertainty were not considered in this work, such as errors in the measurements of redshift, overdensities near emission, and peculiar velocities. The last could be important only at low redshifts, where the population of mergers is of little relevance.

\subsection{Posterior distribution}

We now present our conditional and posterior probabilities employed later on to, first, simulate from a chosen model $\bm{\Theta}$ several synthetic datasets $\bm{d} = (\bm{z}, \bm{D^{\,gw}_L})$, each with $N_{\text{obs}} = 1000$ observation events, and then, to put constraints on the parameters from the generated data, by building the distribution of points that maximize the posteriors (or likelihoods).

\begin{figure}
	\centering
	\includegraphics[scale=0.55]{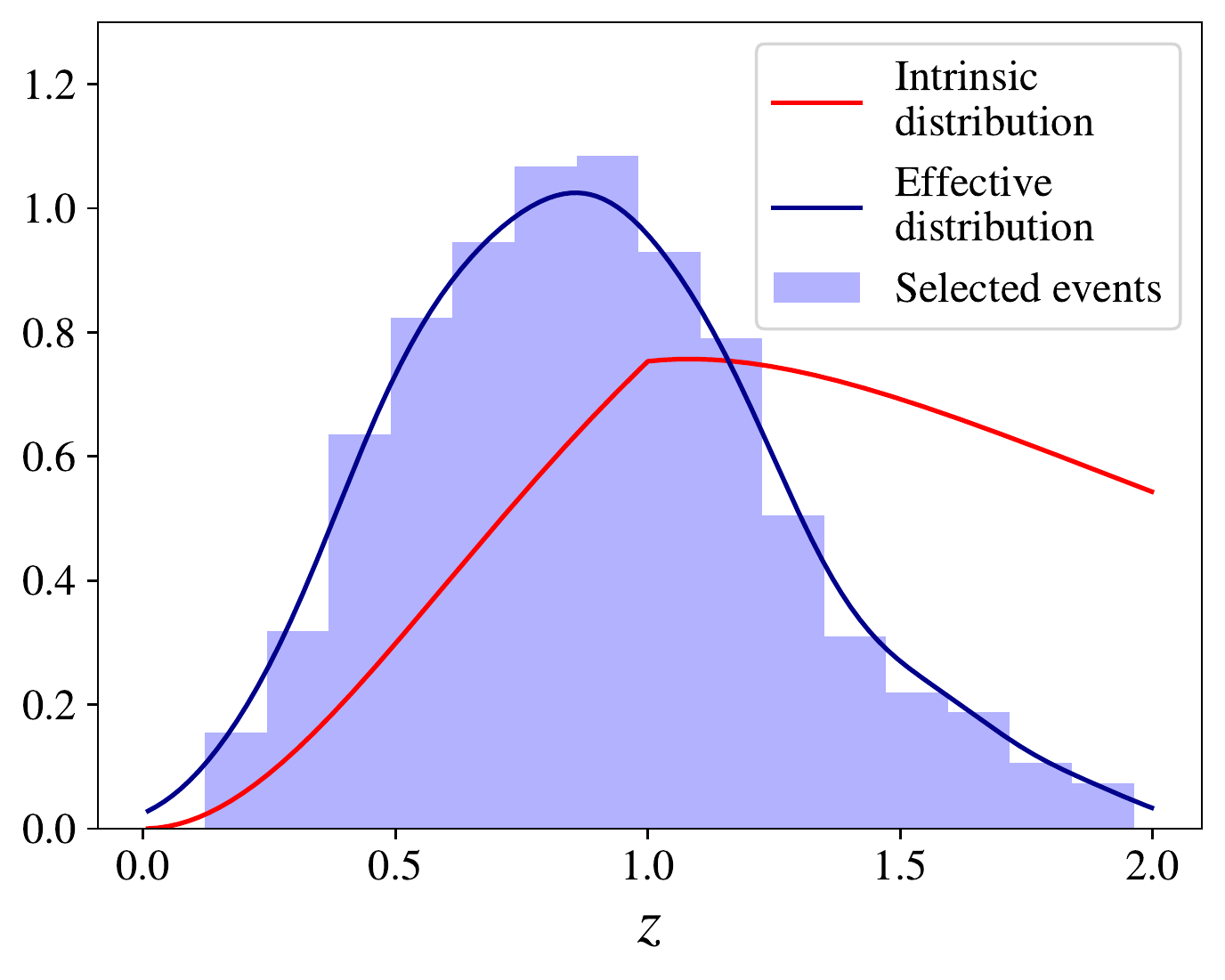}
	\caption{Expected intrinsic [Eq.~(\ref{z_pdf})] and observed or effective distributions of binary neutron star mergers via their gravitational wave signal for the Einstein Telescope. The simulation was made for the $\Lambda$CDM fiducial model.}
	\label{fig:redshift_dist}
\end{figure}

Given a gravitational and cosmological model with parameter values $\bm{\Theta}$, we draw the redshift of the ith binary source from the distribution
\begin{align}
\rho (z_i| \bm{\Theta}) = N_{z}(\bm{\Theta}) \frac{4\pi [\mathcal{D}_c(z_i, \bm{\Theta})]^2}{(1 + z_i) H(z_i, \bm{\Theta})} r(z_i)\,, \label{z_pdf}
\end{align}
where $H$ is the Hubble parameter of the model, $\mathcal{D}_c(z, \bm{\Theta}) \leqdef \int_0^z c/H(z, \bm{\Theta})$ is the comoving distance and $N_z$ is a normalization factor. The rate of BNS mergers evolution, associated to the star formation history itself \cite{Schneider2001}, can be well approximated by
\begin{equation}
r(z) =
\begin{cases}
1 + 2z, & z \leq 1 \\
(15 - 3z)/4, & 1 < z < 5 \\
0, & z \geq 5\,.
\end{cases} \label{burst_rate}
\end{equation}
Of course, the practical probability density of redshifts will be deformed relative to the above function, since we have to discard the unreliable low SNR events (here, $< 8$). In Fig. \ref{fig:redshift_dist} we compare such distributions.

The GW luminosity distance is drawn from a Gaussian distribution around its theoretical value,
\begin{align}
& \rho (D^{\,gw}_{L,i}| \bm{\Theta}, \bm{s}_i, z_i)  = \nonumber \\  & \frac{1}{\sqrt{2\pi}\sigma_i(z_i, \bm{s}_i)} \exp\Bigg\{-\frac{\left[D^{\,gw}_{L,i} - \mathcal{D}^{\,gw}_L(z_i, \bm{\Theta})\right]^2}{2\sigma_i(z_i, \bm{s}_i)^2}\Bigg\}\,.  \label{dl_prob}
\end{align}

Finally, for an arbitrary simulated dataset, one may obtain (similarly to \cite{DAgostini2005}) the posterior probability for the parameters of interest, given the data $\bm{d}$ and source estimates. That is
\begin{align}
\rho(\bm{\Theta}| \bm{z}, \bm{D}^{\,gw}_{L}, \bm{s} ) \propto \rho (\bm{\Theta}) \prod_{i = 1}^{N_{\text{obs}}} \rho (\bm{s}_i) \frac{\mathcal{D}_c(z_i, \bm{\Theta})^2 r(z_i)}{(1 + z_i) H(z_i, \bm{\Theta})} & \nonumber \\ \times \frac{N_z(\bm{\Theta})}{\sigma(z_i, \bm{s}_i)} \exp \Bigg\{-\frac{\left[D^{\,gw}_{L,i} - \mathcal{D}^{\,gw}_L(z_i, \bm{\Theta})\right]^2}{2\sigma(z_i, \bm{s}_i)^2}\Bigg\}\,. & \label{posterior}
\end{align}
The priors $\rho(\bm{s})$ used in this work are presented in Table \ref{tab:variables}. Our procedure consists on finding the values $\bm{\widehat \Theta}$ that maximize this function for each $\bm{d}$ and then build $\rho(\bm{\widehat \Theta})$.

We comment that, in many works such as \cite{DAgostino2019, Belgacem2018a}, only the exponential term and the prior $\rho(\bm{\Theta})$ are considered in the posterior, \textit{i.e.}, at the stage of parameter estimation after the mock data generation, the cosmological and gravitational model is set to the fiducial one in the redshift distribution, as if star formation rate was not sensitive to gravity or the cosmological parameters. In effect, such simplification should be a reasonable approximation in view of the exponential amplification of the difference in distances, so it will be made, hereafter,
\begin{align}
&\rho(\bm{\Theta}| \bm{z}, \bm{D}^{\,gw}_{L}, \bm{s} ) \propto  \nonumber \\ & \hspace{25pt} \rho (\bm{\Theta}) \exp \Bigg\{-\sum_{i = 1}^{N_{\text{obs}}} \frac{\left[D^{\,gw}_{L,i} - \mathcal{D}^{\,gw}_L(z_i, \bm{\Theta})\right]^2}{2\sigma(z_i, \bm{s}_i)^2}\Bigg\}\,. \label{final_posterior}
\end{align}
One should not think, however, that this implies the posterior to be independent of the background evolution and only care about the propagation of tensor modes, \textit{i.e.}, the ratio of luminosity distances, since $\mathcal{D}^{\,gw}_L(z_i, \bm{\Theta})$ itself depends on the EM luminosity distance of the models under inspection.

\section{$\gamma$ gravity} \label{sec:gamma_gravity}

In this section, we briefly discuss spatially flat cosmological models in the context of the so-called $\gamma$ gravity \cite{ODwyer2013}, a viable $f(R)$ theory that generalizes several interesting cases, described by the following expression:
\begin{equation}
f(R) = -\frac{\alpha R_*}{n}\gamma\big(1/n, \left(R/R_*\right)^n\big),
\label{fRgamma}
\end{equation}
where $\gamma(n,x) = \int_0^x t^{n-1} e^{-t}dt$ is the lower incomplete $\Gamma$ function, and $\alpha$, $n$, and $R_*$ are free positive parameters. It follows from Eq.~(\ref{fRgamma}) that the derivatives of  $f(R)$ with respect to $R$ are given by
\begin{align}
f_R & \leqdef \frac{d f}{{d}R} = -\alpha e^{-(R/R_{\ast})^n}, \label{DR} \\
f_{RR} & \leqdef \frac{d^{2}f}{{d}
	R^{2}}= \frac{n \alpha}{R_{\ast}} e^{-(R/R_{\ast})^n}\left(\frac{R}{R_{\ast}}\right)^{n-1}. \label{DRR}
\end{align}
Notice from Eq.~(\ref{DR}) that as we increase $n$, the steepness of the $f(R)$ function increases.
	
It can be shown that $\gamma$ gravity can satisfy all the stability and viability conditions  \cite{Pogosian2008}: (a) $f_{RR} >0$ (no tachyons); (b) $1+f_{R}>0$ [the effective gravitational constant $G_{\text{eff}} = G/(1+f_R)$ does not change sign (no ghosts)]; (c) $\lim_{R\rightarrow \infty }f /R=0$ and $\lim_{R\rightarrow \infty }f _{R}=0$ (GR is recovered at early times); and (d) $|f_R|$ is small at recent epochs (to satisfy solar and galactic scale constraints). Furthermore, Eq.~(\ref{fRgamma}) can also satisfy cosmological viability criteria \cite{Amendola2007}.  We characterize a viable cosmological model as one that starts at a radiation-dominated phase and has a saddle-point matter-dominated phase followed by an accelerated expansion as a final attractor. 
	
Each model is characterized by fixed values of the parameters $\alpha$, $n$, and $R_{\ast}$. Although there is no cosmological constant, $f(0)=0$, at high curvature, when $R \gg R_{\ast}$, the models behave like $\Lambda$CDM. Therefore, by using that $\lim_{x\rightarrow \infty}\gamma\left(\frac1n, x \right)=\Gamma (1/n)$,  from Eq.~(\ref{fRgamma}), at this limit, we get $2\tilde{\Lambda}=\alpha R_{\ast} \Gamma(1/n)/n$, and we can write $R_{\ast}$ as
\begin{equation}
R_* =\frac{6 m^2d} {\alpha\Gamma(1/n)}, \label{Rstar}
\end{equation}
where $d \leqdef (1-\tilde{\Omega}_{m0})/\tilde{\Omega}_{m0}$, and $m^2\leqdef\frac{8\pi G}{3} \bar{\rho}_{m0}$. Here, $\tilde{\Omega}_{m0}$ denotes the present value of the matter density parameter that a $\Lambda$CDM model would have if it had the same matter density today ($\bar{\rho}_{m0}$) as the modified gravity $f(R)$ model. Therefore, if $\tilde{H}_0$ represents the Hubble constant in the reference $\Lambda$CDM model, we have $m^2 = \tilde{\Omega}_{m0}\tilde{H}_0^2=\Omega_{m0}H_0^2$, where $\Omega_{m0}$ and $H_0$ are, respectively, the present value of the matter energy density parameter and Hubble parameter in the $f(R)$ model. Also, since we are mainly interested in $z \leq 2$, we are neglecting radiation in our analysis.
	
To compute the background evolution, we start from the $f(R)$ field equation for a FLRW metric,
\begin{equation}
H^2 (1 + f_R + R_y f_{RR}) - \frac{Rf_R-f}{6} = m^2 e^{-3y}, \label{einstein_eq}
\end{equation}
where $R=12 H^2 + 6 H H_y$ and $y \leqdef \ln a$.
Here, the subindex $y$ stands for the derivative with respect to $y$, $H \leqdef \dot{a}/a$ is the Hubble parameter, and a dot denotes the derivative with respect to the cosmic time.
	
To solve these equations, we introduce the new variables
\begin{align}
x_1 (y) & \leqdef \frac{H^2}{m^2} - e^{-3y} - d,\\
x_2 (y) & \leqdef \frac{R}{m^2} - 3 e^{-3y} - 12\left(d + x_1\right).
\end{align}
With the definitions above we get
\begin{align}
\frac{dx_1}{dy} & = \frac{x_2}{3}, \\
\frac{dx_2}{dy} & = \frac{R_y}{m^2} + 9 e^{-3y} - 4 x_2,
\end{align}
where $R_y$ is given by Eq.~(\ref{einstein_eq}).
	
It is straightforward to verify that, as defined,  $x_1$ and $x_2$ are always zero during the $\Lambda$CDM phase.
Furthermore, in terms of these functions, the effective dark energy equation of state ($w_{\textrm{DE}}$) is given by
\begin{equation}
w_{\textrm{DE}} = -1 - \frac{1}{ 9} \frac{x_2}{x_1 + d}\,.
\end{equation}

Finally, regarding the tensor sector, one may compute the ratio of luminosity distances in Eq.~(\ref{ratio_dist_f(R)}); above $z = 1$, for various values of the parameters $(\alpha, n)$, it is essentially constant and less than 1, since $f_{R0}$ is negative. This behavior is quite common in $f(R)$ gravity theories, which allows parametrizations with 2 degrees of freedom: the asymptote and how fast it goes to it, as we shall discuss next.

\section{Proposal of new parametrizations} \label{sec:parametrizations}

\subsection{Dark energy equation of state parameter}

Recent investigations, based on several cosmological observations, indicate an oscillatory behavior for the dark energy equation of state parameter, $w_{\text{DE}}$ \cite{Zhao2017}.  This kind of behavior with  $w_{\text{DE}}$ crossing the phantom divide line is typical in viable $f(R)$ theories \cite{Hu2007, Starobinsky2007, ODwyer2013}. More recently, \cite{Jaime2018} presented a parametrization for $w_{\text{DE}}$ with four parameters that can reproduce a variety of $f(R)$ models within a $0.5\%-0.8\%$ precision.
Here, also intending to contemplate the typical behavior of the FLRW background in $f(R)$ viable theories, easing numerical computation, we propose a new parametrization for the dark energy equation of state parameter that is a function of three (background) parameters $(A, z_t, z_f)$. It is given by, for $z_f < z < z_t$,
\begin{subequations} \label{wde}
\begin{align}
& w_{\textrm{DE}}(z, A, z_t, z_f) \leqdef -1 \nonumber \\ & - A (z-z_f) (z_t-z) \sin\bigg[\frac{2 \pi z- \pi (z_f+z_t)}{z_t-z_f}\bigg]\,,
\end{align}
and, for $z < z_f$ or $z > z_t$,
\begin{align}
w_{\textrm{DE}}(z,  A, z_t, z_f) \leqdef -1\,.
\end{align}
\end{subequations}

\begin{figure*}
	\centering
	\includegraphics[scale=0.5]{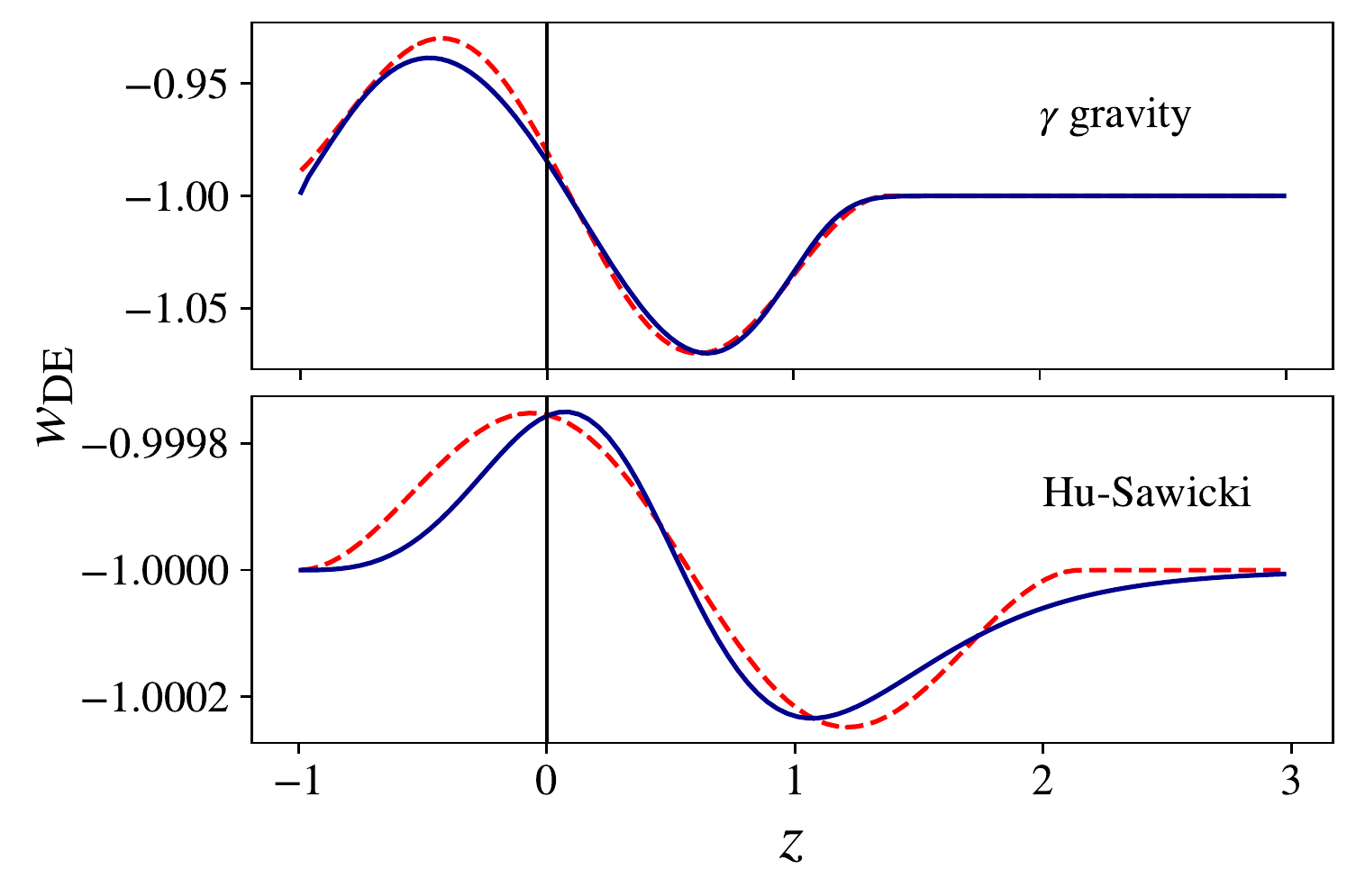}
	\hspace{10pt}
	\includegraphics[scale=0.5]{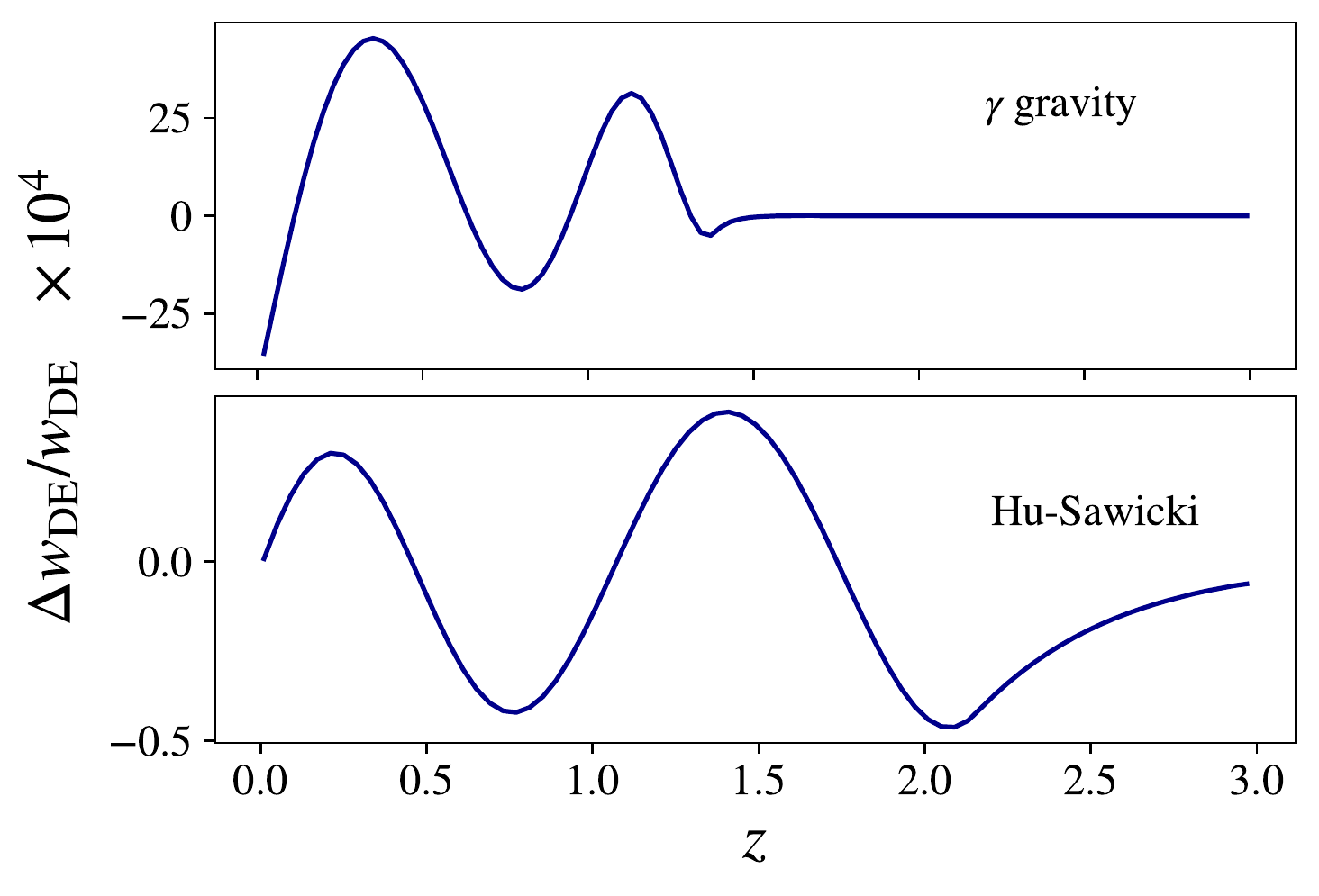}
	\caption{On the left panel, we show a comparison between theoretical predictions (blue, solid) for the dark energy equation of state parameter $w_{\textrm{DE}}$ and the proposed parametrization (red, dashed) fitted for two $f(R)$ theories: the Hu-Sawicki model with $(f_{R0}, n) = (-10^{-4}, 4)$ and $\gamma$ gravity with $(\alpha, n) = (0.9, 2)$. On the right, we see the relative difference between theoretical and phenomenological curves in the past.}
	\label{fig:w_DE}
\end{figure*}

The role of each parameter is immediate: $z_t$ and $z_f$ specify when the oscillating deviation from $w_{\text{DE}}=-1$ starts and ends, respectively, and $A$ controls its amplitude. Generically, we assume $z_t>0$ and $z_f<0$. We remark that, although redshifts smaller than $-1$ are unphysical, here, our main interest is to fit the redshift behavior of the dark energy equation of state only for $z>0$, not in the future ($-1\leq z<0$). We observe, however, that sometimes $z_f <-1$  helps in fitting its desired behavior in the past ($z>0$) and, therefore, we relax the constraint $z_f\geq-1$. Figure \ref{fig:w_DE} shows its fitness for particular $\gamma$ gravity and Hu-Sawicki models.

We find the following advantages of the above parametrization as compared to the one suggested by \cite{Jaime2018}. First, it can also  reproduce the behavior of $w_{\text{DE}}$ in $f(R)$ models within $\sim 0.5\%$ precision (see Fig.~\ref{fig:w_DE}) but it depends on only three (instead of four) parameters. As remarked above, the physical meaning of each parameter is clear. Further, our proposal has a great computational advantage over solving the modified background equations since it allows integrating $3(1 + w_{\text{DE}})/(1 + z)$ to compute the Hubble parameter analytically,
\begin{align}
& H(z, H_0, \Omega_{m0},  A, z_t, z_f)^2  = H_0^2\bigg\{ \Omega_{m0}(1 + z)^3 + \nonumber \\ & (1 - \Omega_{m0}) \exp{\int_0^z\frac{3[1 + w_{\text{DE}}(x,  A, z_t, z_f)]}{1 + x}dx}  \bigg\}\,.
\end{align}

We remark that near the $\Lambda$CDM background, \textit{i.e.}, for $A \rightarrow 0$, $w_{\text{DE}}$ becomes degenerate, since $z_t$ and $z_f$ are arbitrary. To avoid this issue, in this work, we will always keep these two parameters fixed.


\subsection{Ratio of luminosity distances}

The test we are interested in this work involves the ratio $\mathcal{D}_L^{\,gw}(z)/\mathcal{D}_L^{\,em}(z)$. Instead of calculating this quantity for each $f(R)$ model, it is convenient to consider a simple parametrization able to describe the main features of several models. The parametrization we propose here is defined by
\begin{align}
\Xi(z, \Xi_0, \nu) \leqdef \Xi_0 + (1-\Xi_0) e^{1 - (1 + z)^{\nu}}\,, \label{us}
\end{align}
so that
\begin{align}
\mathcal{D}^{\,gw}_L(z, \bm{\Theta}) = \Xi(z, \Xi_0, \nu) \mathcal{D}^{\,em}_L(z, H_0, \Omega_{m0}, A, z_t, z_f)\, \label{dl_parametrized}
\end{align}
should be replaced in the posterior (\ref{final_posterior}). The GW luminosity distance thereby becomes a function of redshift, cosmological and background parameters, and the now-introduced two parameters $ (\Xi_0, \nu)$ associated to the tensor modes evolution.

\begin{figure*}[t]
	\centering
	\includegraphics[scale=0.5]{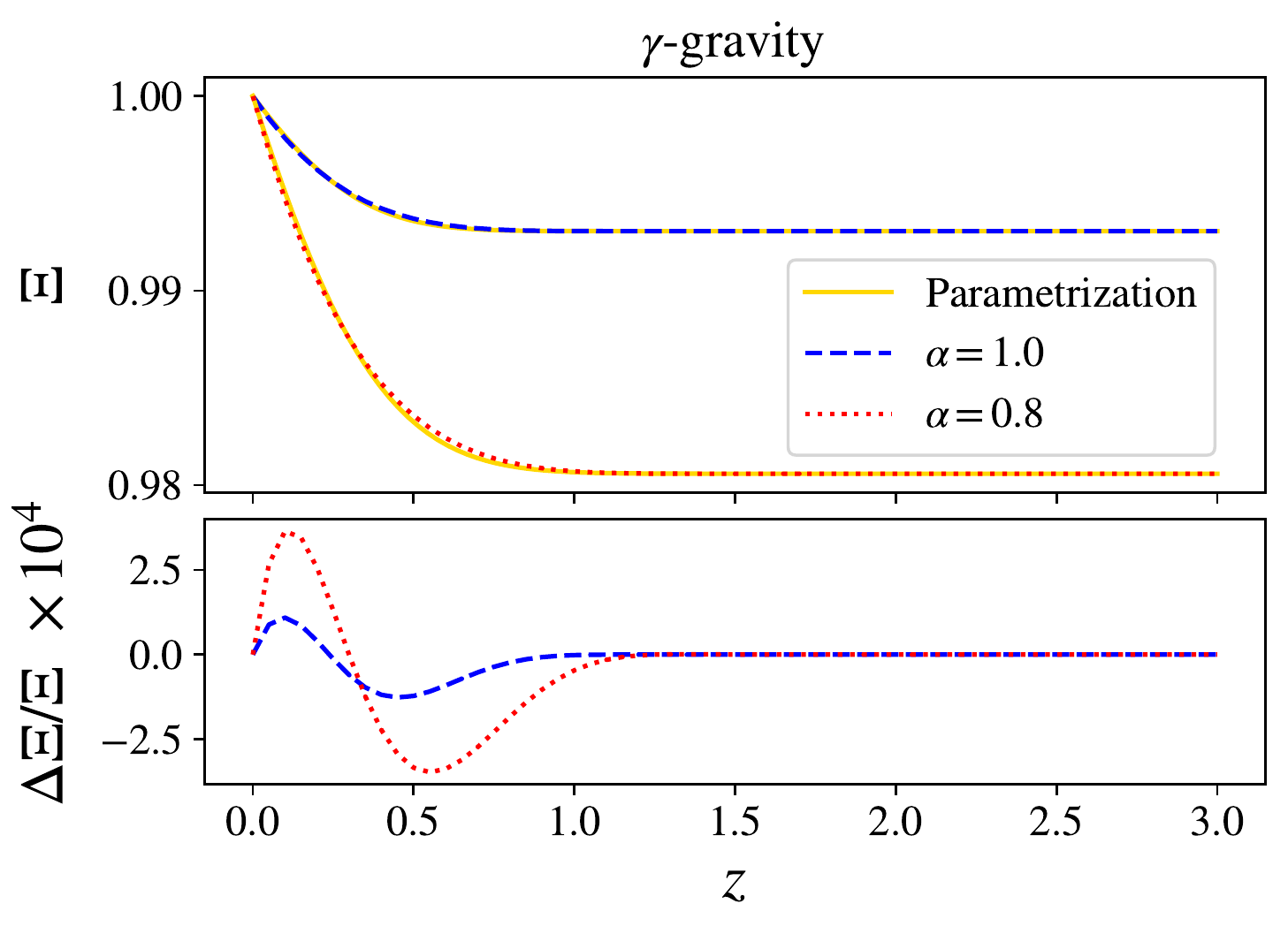}
	\hspace{10pt}
	\includegraphics[scale=0.5]{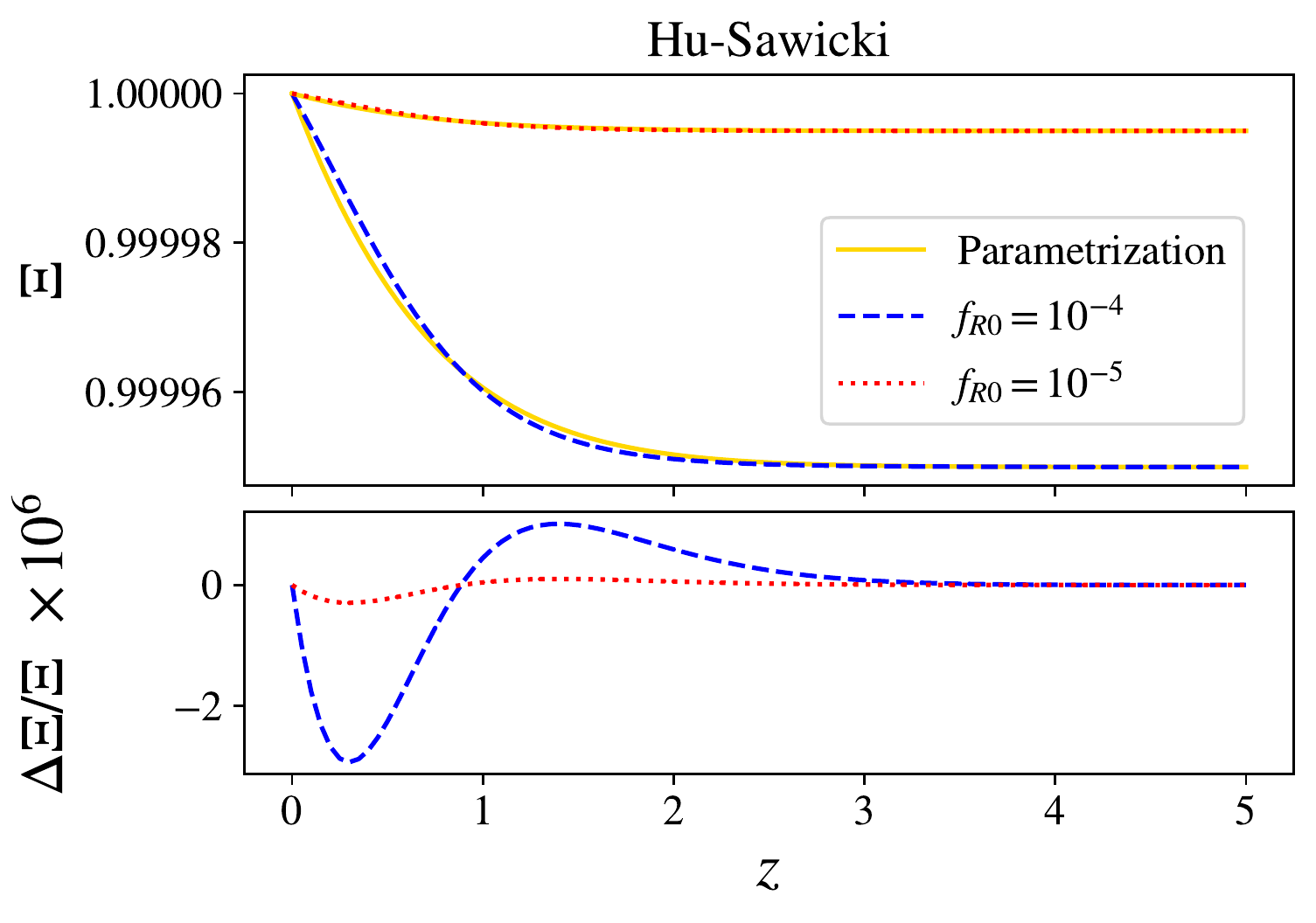}
	\caption{Comparison between theoretical predictions for the ratio of luminosity distances with the fitted proposed parametrization in (left) $\gamma$ gravity and (right) Hu-Sawicki, both for $n = 2$. The lower panels show the relative difference between the dashed or dotted curves and the corresponding solid yellow one.}
	\label{fig:ratio_lum_dist}
\end{figure*}

This parametrization has the same spirit as the one considered in \cite{Belgacem2018a} (see also \cite{Belgacem2019}),
\begin{align}
\Xi(z, \Xi_0, \nu) =\Xi_0 + \frac{(1-\Xi_0)}{(1 + z)^{\nu}}, \label{belga}
\end{align}
in the sense that it is equal to unity at $z=0$ and goes to a constant $\Xi_0 $ for high redshift. Both expressions are degenerate to the GR value ($\Xi=1$) at either $\Xi_0=1$ or $\nu=0$. In fact, an identical relation holds for Eqs.~(\ref{us}) and (\ref{belga}) when imposing that the total variation of $\Xi$ vanishes, namely
\begin{align}
\frac{\partial \Xi}{\partial \Xi_0}\delta\Xi_0 = -\frac{\partial \Xi}{\partial \nu}\delta \nu \;\; \Rightarrow \;\; \left|\frac{\delta \nu}{\nu}\right| = \left|\frac{\delta\Xi_0}{1 - \Xi_0}\right|\,,
\end{align}
and thus, the uncertainty in $\nu$ ($\Xi_0$) diverges when $\Xi_0 \rightarrow 1$ ($\nu \rightarrow 0$). We noticed that Eq.~(\ref{belga}) fits better the ratio of distances in nonlocal models, while Eq.~(\ref{us}), which will be used throughout this work, describes better its behavior in the case of $f(R)$, as illustrated in Fig.~(\ref{fig:ratio_lum_dist}), which shows its fitness to $\gamma$-gravity models.

\section{Einstein Telescope forecasts} \label{sec:results}

Here, we finally present our Einstein Telescope forecasts from several synthetic datasets generated by the fiducial model being either $\Lambda$CDM or $\gamma$ gravity with $n = 2$ and $\alpha = 0.9$. We held cosmological and gravitational (phenomenological or fundamental) parameter estimation and looked at whether it is possible to distinguish these models from each other. Table \ref{tab:results} summarize our results.

\begin{table*}[t] 
	\centering
	\caption{Summary of results}
	\label{tab:results}
	\setlength{\extrarowheight}{8pt}
	\begin{tabular}{|c|c|c|c|}
		\hline
		Version & 1D 68\% C. Ls. & Priors & Fiducial model \\
		\hline
		(1) & $\Delta H_0 = \pm 0.69$, $\Delta \Omega_{m0} = ^{+0.023}_{-0.026}$ & No priors  & \multirow{7}{*}{$\Lambda$CDM}\\ \cline{1-3}
		(2) & $\Delta H_0 = \pm 0.3$, $\Delta \Omega_{m0} = \pm 0.002$ 	& Gaussian with $\sigma_{H_0} = 1.9$, $\sigma_{\Omega_{m0}} = 0.007$ & \\ \cline{1-3}
		(3) & $\Delta f_{R0} = \pm 0.015$ & Flat in [-1, 1] &  \\ \cline{1-3}
		\multirow{2}{*}{(4)} & $\Delta H_0 = \pm 0.82$, $\Delta \Omega_{m0} = ^{+8.4}_{-6.8} \times 10^{-4}$, & Gaussian with $\sigma_{H_0} = 1.9$, $\sigma_{\Omega_{m0}} = 0.007$, &  \\
		& $\Delta A = \pm 0.069$, $\Delta \Xi_0 = \pm 0.016$ & $\sigma_A = 0.14$; fixed $\nu = 2.82$ &  \\ \cline{1-3}
		\multirow{2}{*}{(5)}  & $\Delta \Omega_{m0} = ^{+0.10}_{-0.15}$, $\Delta A = ^{+0.13}_{-1.00}$ & No priors &  \\ \cline{2-2}
		& $\Delta \Omega_{m0} = \pm 0.012$, $\Delta A = ^{+0.13}_{-0.12}$ & No priors \footnotesize{(GW+SN+BAO/CMB)}	  &  \\ \cline{1-4}
		\multirow{2}{*}{(6)} & \multirow{2}{*}{$\Delta A = ^{+0.063}_{-0.047}$, $\Delta \Xi_0 = \pm 0.014$}  & Gaussian with $\sigma_{H_0} = 1.9$, $\sigma_{\Omega_{m0}} = 0.007$, & \multirow{4}{*}{$\gamma$-gravity} \\ 
		& & $\sigma_A = 0.14$; fixed $\nu = 2.82$  & \\ \cline{1-3}
		\multirow{2}{*}{(7)} & \multirow{2}{*}{$\Delta A = \pm 0.057$, $\Delta \nu = ^{+1.8}_{-2.7}$} & Gaussian with $\sigma_{H_0} = 1.9$, $\sigma_{\Omega_{m0}} = 0.007$, &\\ 
		& & $\sigma_A = 0.14$; fixed $\Xi_0$ & \\
		\hline
\end{tabular}
\end{table*}

\subsection{$\Lambda$CDM}

\begin{figure}[h!]
	\includegraphics[scale=0.5]{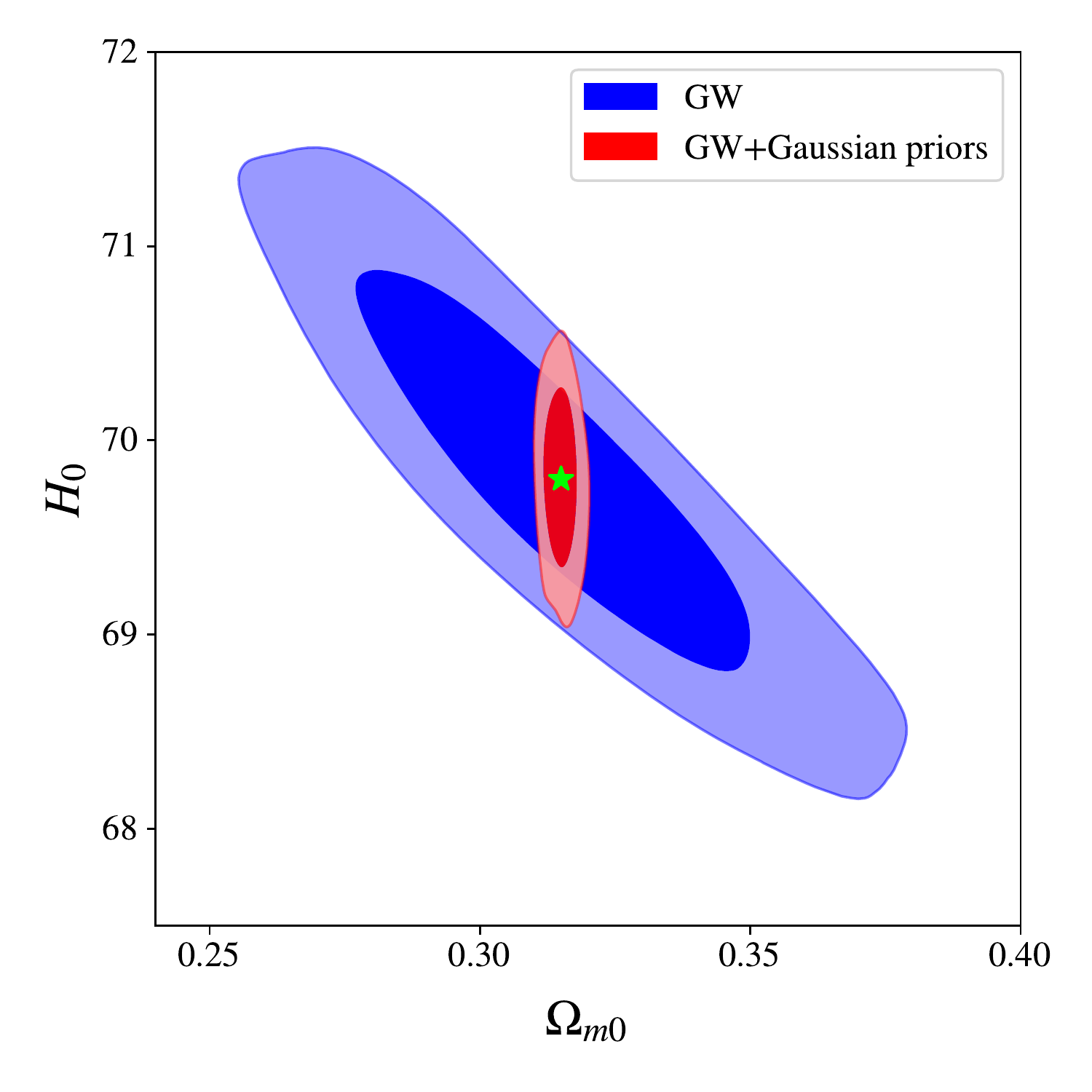}
	\caption{ET forecast for $H_0$ and $\Omega_{m0}$ for GR's $\Lambda$CDM as fiducial model (green star). 2D confidence regions refer to (68\%, 95\%) probabilities.}
	\label{fig:LCDM}
\end{figure}

First, restricting ourselves to spatially flat $\Lambda$CDM and neglecting radiation, the posterior only depends on two parameters,
\begin{align}
	\bm{\Theta} = (H_0, \Omega_{m0})\,,
\end{align}	
for which we have chosen the fiducial values
\begin{align} \label{fid_model_interm_H0}
H_0^{\text{fid}} = 69.8\,, \; \Omega_{m0}^{\text{fid}} = 0.315\,,
\end{align}
from \cite{Freedman2001} and \cite{Lahav2019}, respectively. Figure \ref{fig:LCDM} shows the expected constraints in two cases (versions 1 and 2 of Table \ref{tab:results}): blue when considering the distribution of points $\bm{\widehat{\Theta}}$ maximizing the likelihoods, one best fit for each simulated dataset, and red when imposing Gaussian priors centered in the fiducial model with standard deviations coming from other probes equals $1.9$ \cite{Freedman2001} and $0.007$ \cite{Lahav2019}, respectively.

As regarding the marginalized 68\% levels, we found that GW data from ET only will be able to estimate these parameters with accuracy of
\begin{align}
\frac{\Delta H_0}{H_0} = 1.0\% \,, \hspace{10pt} \frac{\Delta \Omega_{m0}}{\Omega_{m0}} = 7.6\%\,.
\end{align}

\subsection{Constraints to $f(R)$ in the asymptotic regime} \label{subsec:fR0}

Considering the simplifying picture discussed in Sec \ref{sec:GW_in_MG}, \textit{i.e.}, the ratio of luminosity distances as given by Eq.~(\ref{asymptotic_lum_dist}), and that the $\Lambda$CDM background evolution is a good approximation, we simulate 1000 detections by ET of GWs emitted by BNS mergers only at redshifts between 1 and 2. The fiducial model employed in the mock data was GR's $\Lambda$CDM, according to Planck, 2018 \cite{Planck2018}, with $H_0^{\text{fid}} = 67.66$, $\Omega_{m0}^{\text{fid}} = 0.310$, and, of course, $f_{R0}^{\text{fid}} = 0$. These three are the parameters $\bm{\Theta}$ upon which our posterior depends in this case. Fixing the first two thenceforth, the posterior,
\begin{align}
\rho (f_{R0}|\bm{d}, \bm{s}) \propto \exp\left\{\sum_{i = 1}^{N_{\text{obs}}}\frac{\left[D_{L,i}^{\,gw} - \mathcal{D}_L^{\,gw}(z_i, f_{R0})\right]^2}{2\sigma_i^2}\right\}\,,
\end{align}
which is now Gaussian due to the linear dependence in Eq.~(\ref{asymptotic_lum_dist}), can be exactly maximized for each simulated dataset $\bm{d}$, resulting in
\begin{subequations}
\begin{align}
\hat{f}_{R0} &= \sigma_{f_{R0}}^2 \sum_{i = 1}^{N_{\text{obs}}} \frac{\left[D^{\,gw}_{L,i} - \mathcal{D}^{\,em}_L(z_i)\right]}{2 \sigma_i^2}\mathcal{D}^{\,em}_L(z_i),\\
& \hspace{10pt} \sigma_{f_{R0}} = \bigg\{\sum_{i = 1}^{N_{\text{obs}}}\bigg[\frac{\mathcal{D}^{\,em}_L(z_i)}{2\, \sigma_i}\bigg]^2 \bigg\}^{-\frac{1}{2}}\,. \label{sigma_fR0}
\end{align} 
\end{subequations}

\begin{figure}
	\centering
	\includegraphics[scale=0.5]{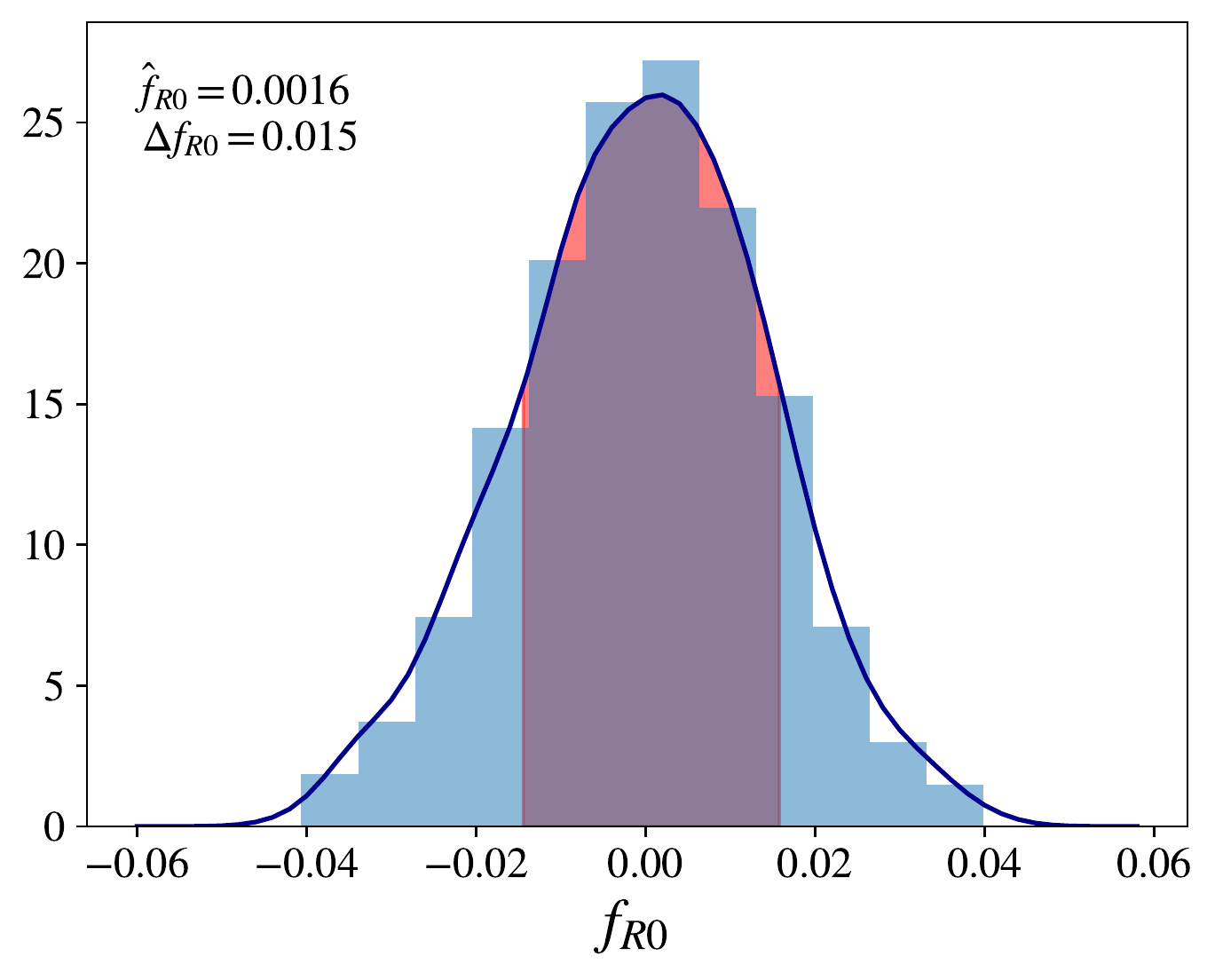}
	\caption{Constraints to $f_{R0}$ from 1000 hypothetical GW detections by ET, valid for any viable $f(R)$ model whose background evolution is nearly GR's $\Lambda$CDM before $z = 1$. The 68\% confidence region of half width $\Delta f_{R0}$ is shown, as well as the peak value $\hat{f}_{R0}$.}
	\label{fig:fR0}
\end{figure}

\begin{figure}
	\includegraphics[scale=0.5]{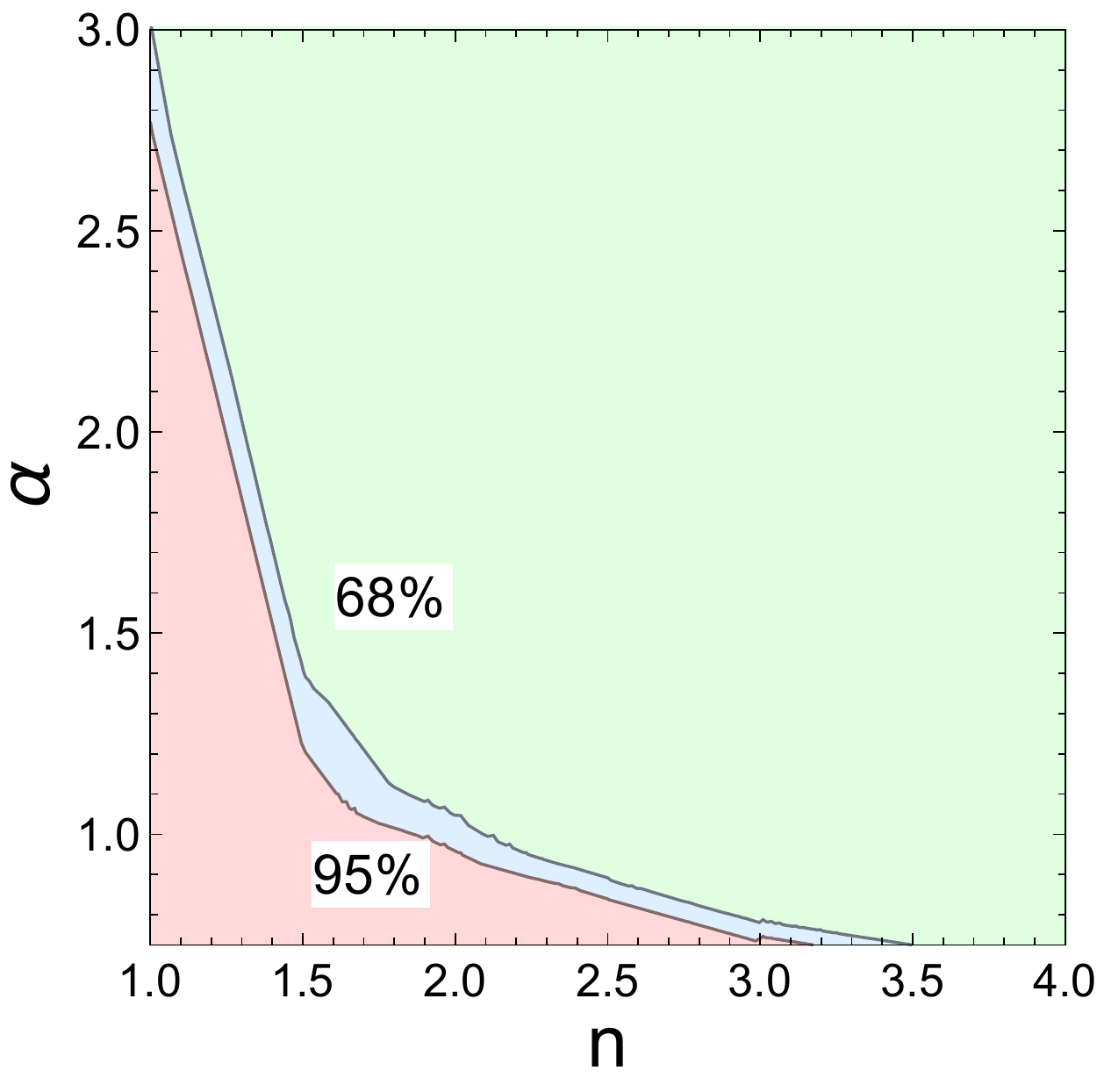}
	\caption{68\% (green) and 95\% (blue and green) confidence regions in the parameter space of $\gamma$ gravity from the hypothetical constraint to $f_{R0}$.}
	\label{fig:alpha_n}
\end{figure}

The distribution $\rho(\hat{f}_{R0})$ is shown in Fig.~\ref{fig:fR0}. Positive values were allowed, but only negative ones correspond to viable models. We conclude that, by looking at GW sources above $z = 1$, for $f(R)$ viable models whose background is nearly $\Lambda$CDM at this epoch, ET alone will be able to distinguish it from the standard model at $95\%$ level only when $|f_{R0}| > 3 \times 10^{-2}$ . For the $\gamma$-gravity parameter space, this constraint translates to what is shown in Fig.~(\ref{fig:alpha_n}). Note that one has to only look at the subspace for which the simplifying hypothesis hold.

Furthermore, a rough relation between the uncertainties of $D_L^{\,gw}$ and $f_{R0}$ can be obtained by looking at Eqs.~(\ref{sigma_fR0}) and (\ref{sigma}). The lower bound for the relative error $\sigma_i/\mathcal{D}_L^{\,gw}(z_i, f_{R0})$ is of about 5\% considering that for very high SNR, the lensing error dominates, and its minimum occurs at the lowest redshift considered ($z = 1$). An upper bound of 25\% is settled by the SNR cutoff. Thus, in the most optimistic case,
\begin{align}
\sigma_{f_{R0}} = & \frac{(2 + f_{R0})}{\sqrt{N_{\text{obs}}}} \bigg<\bigg[\frac{\mathcal{D}^{\,gw}_L(z_i)}{ \sigma_i}\bigg]^2\bigg>^{-\frac{1}{2}}_{\hspace{-4pt}\bm{d}} \nonumber\\
\simeq & \frac{10\%}{\sqrt{N_{\text{obs}}}}\,,
\end{align}
and so, in order to obtain an accuracy of $10^{-6}$ with ET only, it would be necessary to detect the unexpected amount of $N_{\text{obs}} \simeq 10^{10}$ GW signals.

\subsection{Constraints to the phenomenological parameters}

We now look at a larger set of gravitational and cosmological models, taking into account its deviations from GR's $\Lambda$CDM both in the background evolution and in the tensor modes propagation, making use of the discussed parametrizations over the entire range of $0.01 < z < 2$. Therefore, fixing $z_f$ and $z_t$, our parameters of interest become, in this context,
\begin{align}
\bm{\Theta} = (H_0, \Omega_{m0}, A, \Xi_0, \nu)\,,
\end{align}
although, whenever $\Xi_0$ is free, $\nu$ will be fixed and vice versa to avoid degeneracy.

\begin{figure*}[t]
	\includegraphics[scale=0.5]{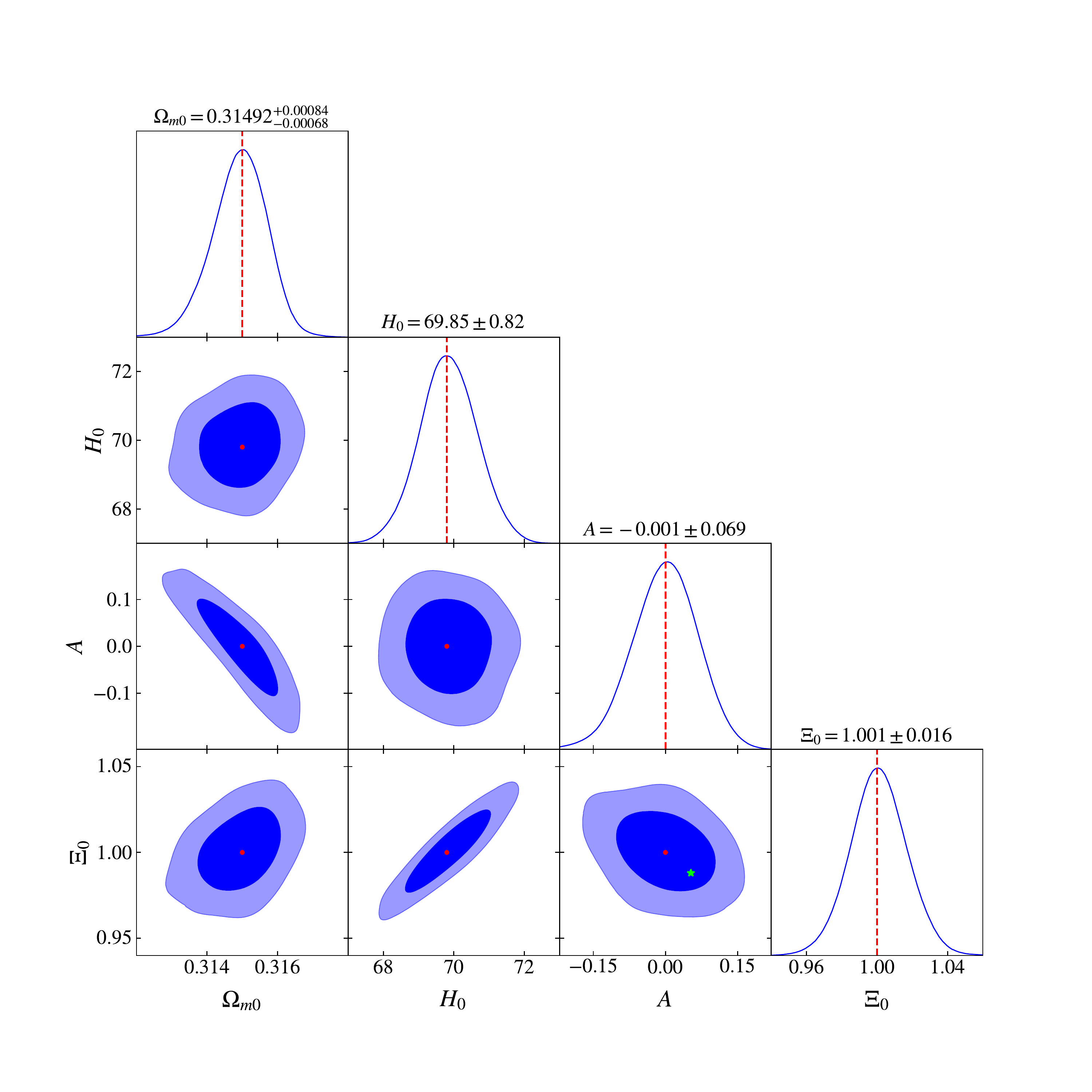}
	\caption{Marginalized 2D confidence regions (68\% and 95\%) obtained from GW simulated data out of $\Lambda$CDM fiducial model (red), relative to version 4 of Table \ref{tab:results}. Titles indicate the marginalized 1D 68\% intervals. The $\gamma$ gravity model with $(n, \alpha) = (2, 0.9)$ is shown for comparison (green star).}
	\label{fig:v12b}
\end{figure*}

First, regarding actual data from other probes, we adapted to our parametrization of $w_{\text{DE}}$ both the SN code used in \cite{Arjona2019} for the Pantheon data \cite{Scolnic2018} and the BAO/CMB analysis of \cite{dosSantos2016b} with data from \cite{Padmanabhan2012, Anderson2012, Blake2011, Beutler2011}. The covariance matrix of the latter was updated according to Planck's 2018 results (TT,TE,EE+lowE+lensing in Table II of \cite{Planck2018}). A combined SN+BAO/CMB marginalized 68\% bound for $A$ of $\Delta A = \pm 0.14$ was then obtained. Additionally to the Gaussian priors in $H_0$ and $\Omega_{m0}$ in the GW posterior, we then used one for $A$ with this value as the standard deviation. With $\Lambda$CDM of Eq.~(\ref{fid_model_interm_H0}) as fiducial model and fixed $\nu$, the results are shown in Fig.~\ref{fig:v12b}.

\begin{figure}[t]
		\includegraphics[scale=0.5]{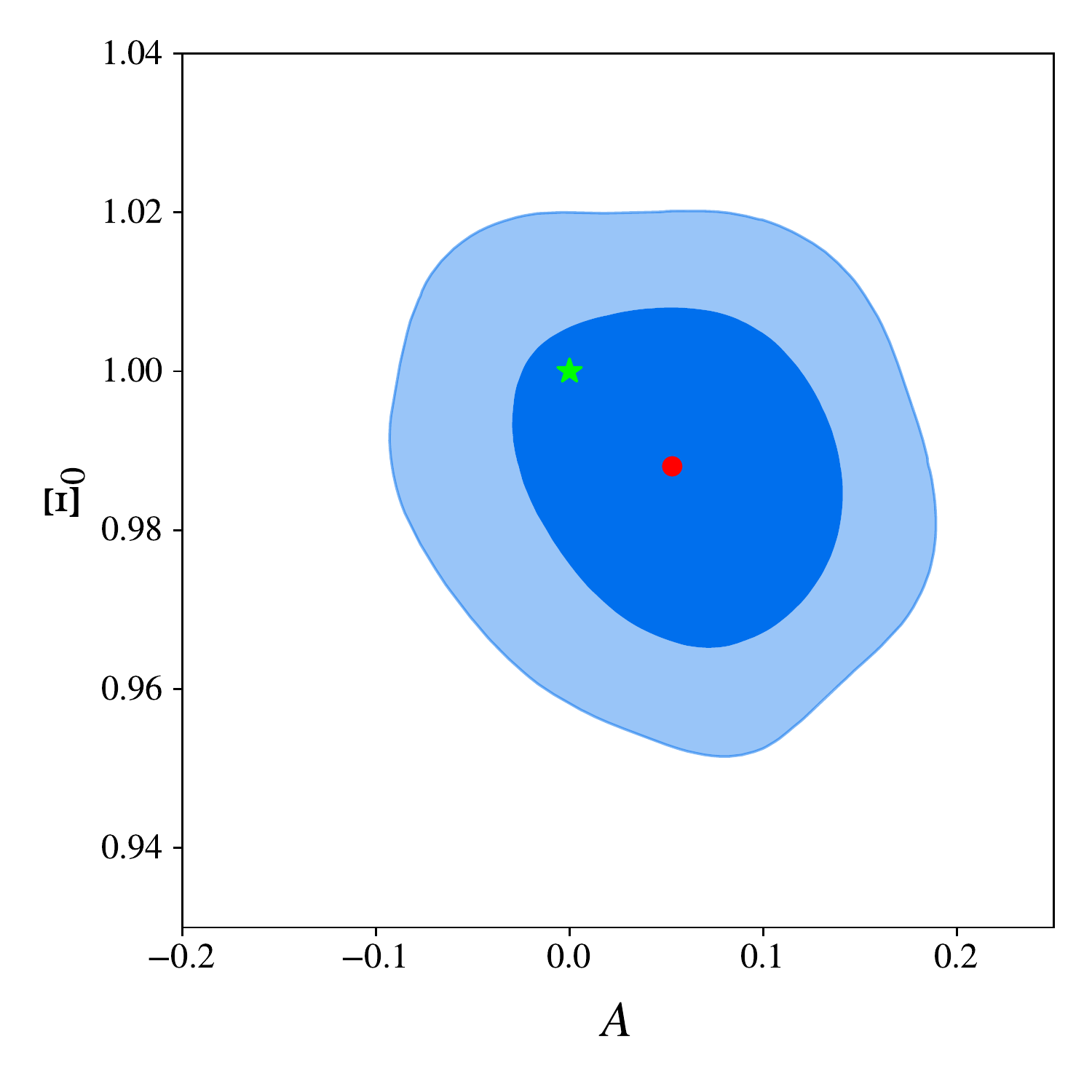}
		\caption{Marginalized 2D confidence regions (68\%, 95\%) from 1000 simulated observations of GW by ET. It corresponds to version 6 of Table \ref{tab:results} with fixed $\nu$. The fiducial model of $\gamma$ gravity (red) and $\Lambda$CDM (green star) are shown for comparison.}
		\label{fig:v10b_v16b}
\end{figure}

In terms of the phenomenological parameters, the $\gamma$-gravity model $(n, \alpha) = (2, 0.9)$, chosen as an example, corresponds to $A = 0.053\,, \;  z_t = 1.38\,, \; z_f = -1.194\,, \; \Xi_0  = 0.988\,,$ and $\nu = 2.82$.
This model purposely has a quite large value for $|f_{R0}|$ of $\sim 10^{-2}$, and thus, it is in the border of the 95\% bound from the simplifying analysis of Sec. \ref{subsec:fR0} (cf. Figs. \ref{fig:fR0}, \ref{fig:alpha_n}). Here, we concluded that it can not be distinguished from $\Lambda$CDM even when considering GWs emitted from the closest sources and accounting for background modifications. In fact, we held similar simulations with this MG model as the fiducial one (and the same Gaussian priors around it), which resulted to be compatible with  $\Lambda$CDM, either by fixing $\nu$ or $\Xi_0$ (versions 6 and 7 of Table \ref{tab:results}), as exemplified in Fig.~\ref{fig:v10b_v16b}.

\begin{figure*}[t]
	\includegraphics[scale=0.5]{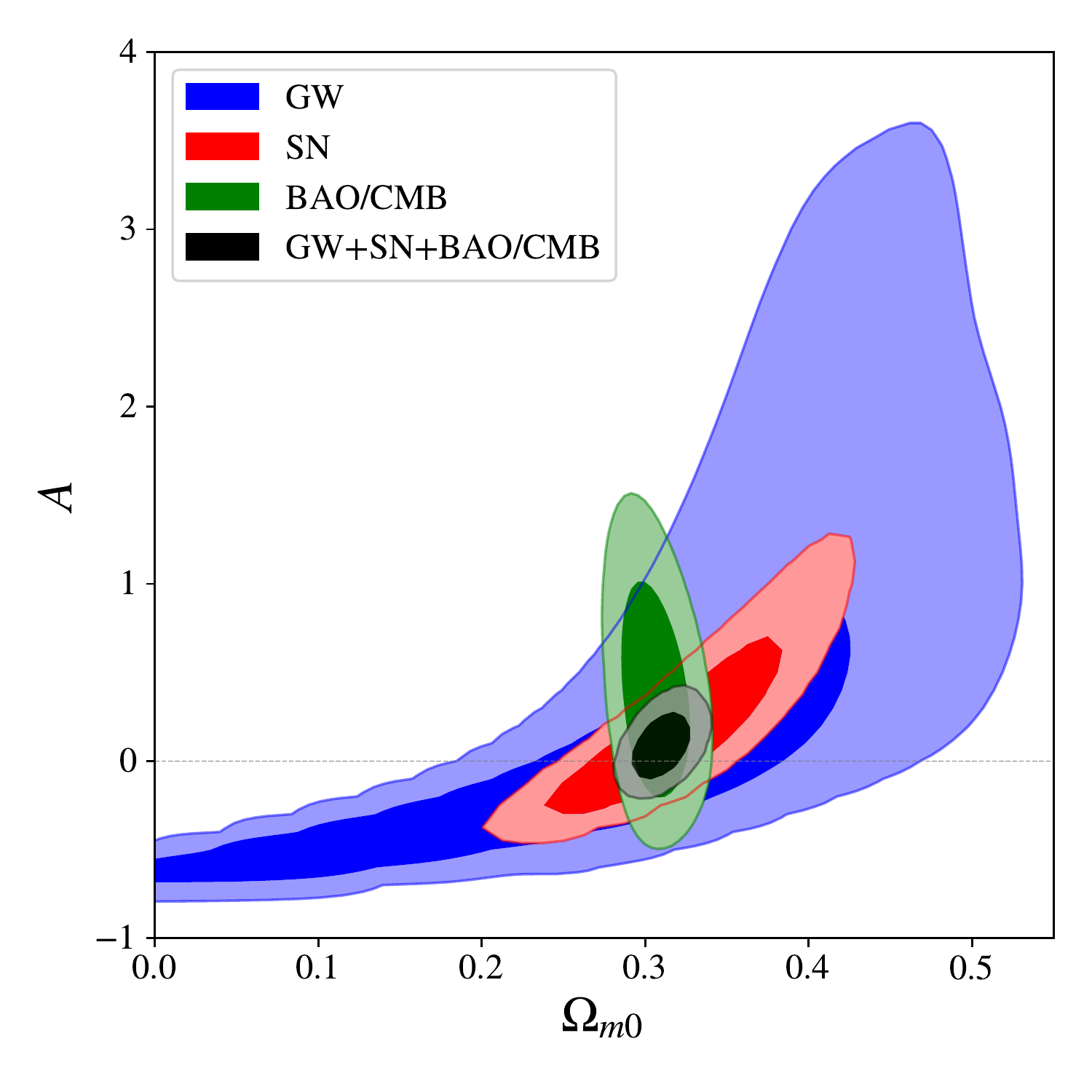}
	\hspace{10pt}
	\includegraphics[scale=0.495]{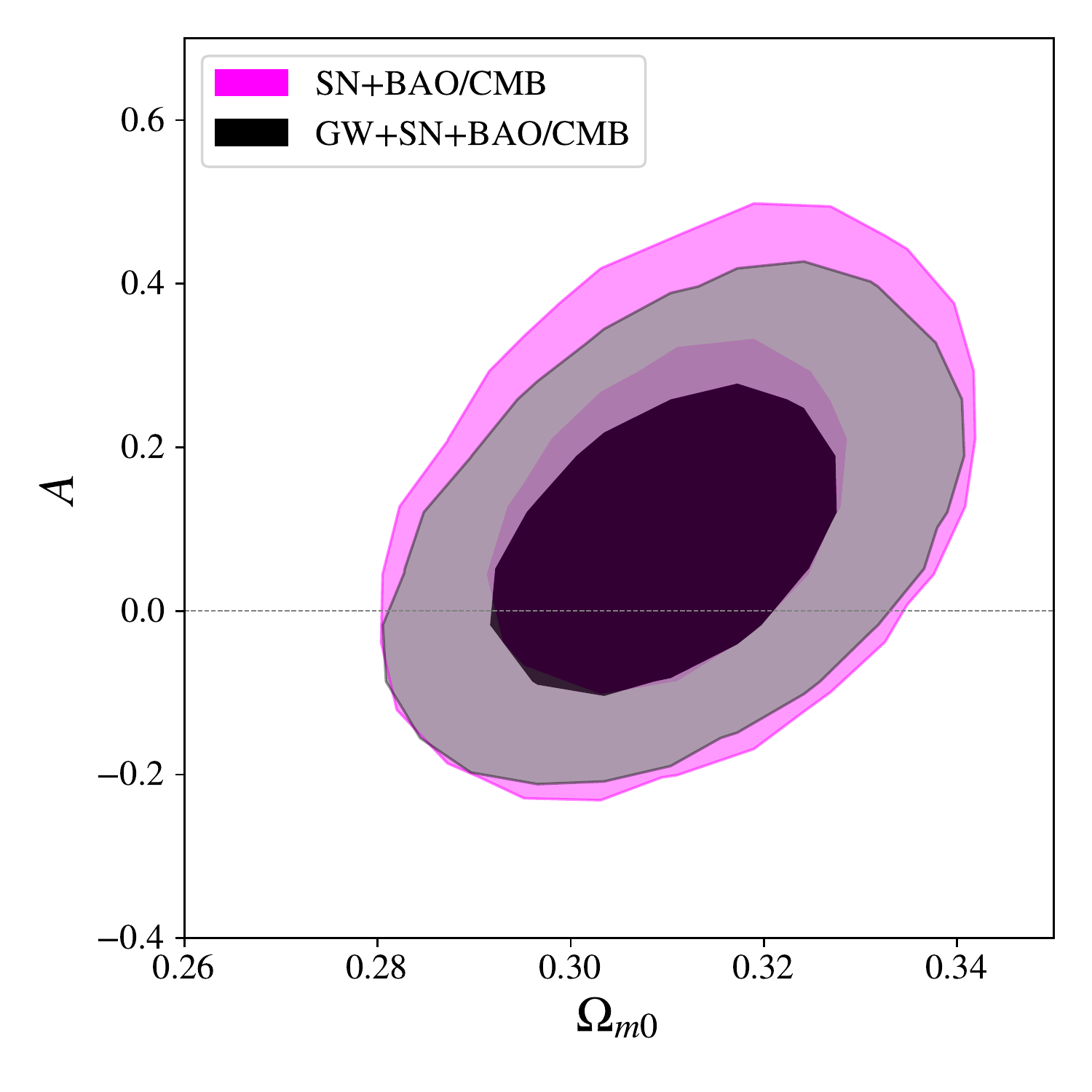}
	\caption{Constraints to $(A, \Omega_{m0})$ from 1000 hypothetical GW detections by ET compared and combined with results from SN and BAO/CMB. It corresponds to version 5 in Table \ref{tab:results}.}
	\label{fig:combined_results}
\end{figure*}

Finally, we also performed maximizations of the GW likelihoods with no priors but fixed $\nu$ and then combined these forecasts with the results of SN and BAO/CMB, as presented in Fig.~\ref{fig:combined_results}. We see from the right panel that a minor improvement is achieved with the inclusion of standard sirens, reducing the 68\% (95\%) area in the $(\Omega_{m0}, A)$ plane in 17\% (18\%).

\section{Discussion} \label{sec:discussion}

In this work, we proposed two independent phenomenological parametrizations in order to describe the evolution of the GW luminosity distance in $f(R)$-like theories of gravity, Eq.~(\ref{wde}) for $w_{\text{DE}}$ and Eq.~(\ref{us}) for $\mathcal{D}^{\,gw}_L/\mathcal{D}^{\,em}_L$, and studied their adequacy to $\gamma$-gravity and Hu-Sawicki models. We then generated a mock dataset of 1000 GWs emitted by BNS mergers as could be detected by the ET in near future, and we investigated the constraints that the inference of the GW luminosity distances and redshifts of the sources could provide to the cosmological and phenomenological parameters. As main results, it was concluded that this kind of test with ET will give an independent 1\% accuracy measurement of the Hubble constant for a $\Lambda$CDM universe. However, regarding $f(R)$ models, first, it was shown to provide weak bounds to $f_{R0}$ when restricting to the farthest sources where the background evolution is nearly $\Lambda$CDM (see Fig.~\ref{fig:fR0}); second, even when considering nearest sources and accounting for the modified background with the parametrization, we showed that it was not possible to distinguish between $\Lambda$CDM and a quite different $f(R)$ MG model, easily discarded with the stringent bounds to $f_{R0}$ from other probes. We also combined our constraints from GWs with actual data from SN and BAO/CMB, and we showed that it does not result in great improvement in the measurement of $\Omega_{m0}$ and the amplitude of the dark energy equation of state parameter deviation from -1. All of it indicates, therefore, that the test is not suitable to look for signatures of $f(R)$ modifications of gravity, neither from dark energy effects or modified friction in GW propagation, even in our, perhaps optimistic, 1000 event scenario.

For a generic family of MG theories, particularly for $f(R)$, given its Lagrangian parametrized by $\bm{l}$, both the FLRW background, scalar, and tensor modes evolution are theoretically determined as functions of cosmological parameters and $\bm{l}$. In most of this work, we treated background (via $w_{\text{DE}}$) and GWs (via $\Xi$) as independent degrees of freedom, and thus, our phenomenological parameter space comprises, in general, a wider set (with larger dimension) of gravity models. Although it is easy to fit the parametrizations to a given model, \textit{i.e.}, to find (numerically) $A, z_t, z_f, \Xi_0, \nu$ as functions of $\bm{l}, H_0, \Omega_{m0}$, this is why the way back is not obvious, making it subtle to translate the constraints obtained for the phenomenological and cosmological parameters to the original parameter space $(\bm{l}, H_0, \Omega_{m0})$ and thus, constrain the family of Lagrangians. Still, through the approach here adopted, it is possible to investigate, for example, whether a given $f(R)$ MG model of interest can be ruled out or distinguished from GR's $\Lambda$CDM by future GW observations of other planned detectors.

Another way of easily computing the ratio of luminosity distances for a whole class of gravity models is by parametrizing $\delta$ in Eq.~(\ref{ratio_dist}) instead of $\Xi$ itself. Although it can be more directly calculated from a given theory, one has to worry about whether its integration does not prejudice the quality of the fit since the actual measured quantity is $D^{\,gw}_L$. In the context of property functions \cite{Bellini2014}, proposals like $-2\delta = \alpha_M = \alpha_{M0}a^s$ \cite{Denissenya2018} have been studied, but this one in particular enjoys a similar degeneracy problem at GR to the one here proposed. An interesting perspective is, therefore, to try out a new parametrization without this problem, and that, at least for $f(R)$ theories, could unify both $w_{\text{DE}}$ and $\Xi$, since their deviations from the $\Lambda$CDM's respective behaviors must occur at the same epoch and with somehow related amplitudes.

In \cite{Belgacem2018a}, the authors claim, in the context of nonlocal theories of gravity, that the contribution for the GW luminosity distance coming from the deviations in the background are less relevant than those coming from the friction term. Here, we comment that, in the case of $f(R)$ theories, these contributions can be, in principle, equally relevant, since the ratio between GW and EM distances is restricted to be a small deviation from 1, in view of the constraints to $f_{R0}$ from other probes. In \cite{DAgostino2019}, besides using a different function $\Xi$, derived from a parametrization for $\delta (z)$, the authors do not take into account modifications to the background when studying the corresponding phenomenological class of MG theories. Also, as examples of differences in approach, we mention that in \cite{Belgacem2018a}, as in \cite{Zhao2011}, the instrumental error in the GW luminosity distance is averaged over the angular variables, besides the fact that these authors run a MCMC to reconstruct a single posterior while we perform maximizations of several distributions. Despite that, our results to the $\Lambda$CDM case (cf. Fig.~\ref{fig:LCDM}) are in reasonable agreement.

\paragraph*{Acknowledgments.} ---
 We thank Dr. Marcelo V. dos Santos for helpful suggestions. I. S. M. thanks Brazilian funding agency CNPq for Ph.D. scholarship GD 140324/2018-6.
 
%
%

\end{document}